\documentclass[12pt]{iopart}

\usepackage{xcolor}
\usepackage{gensymb}
\usepackage{graphicx}
\usepackage{bm}
\usepackage{subfig}
\usepackage{hyperref}
\usepackage{amssymb}
\usepackage{float}
\usepackage{dcolumn}
\usepackage{braket}
\usepackage[normalem]{ulem}
\usepackage{upgreek}
\usepackage{soul}
\usepackage{multirow}
\usepackage{enumitem,kantlipsum}
\usepackage{lineno,hyperref}
\usepackage{lipsum}  
\hypersetup{colorlinks=false}
\usepackage{color}
\usepackage{caption}

\begin{document}
	\title[Enhanced beam shifts mediated by Bound States in Continuum]{Enhanced beam shifts mediated by Bound States in Continuum}
	
	\author{Sounak Sinha Biswas$^1$, Ghanasyam Remesh$^2$, Venu Gopal Achanta$^2$, Ayan Banerjee$^{1,3}$, Nirmalya Ghosh$^{1,3}$ \& Subhasish Dutta Gupta$^{1,4,5\dagger}$}
	
	\address{$^1$Department of Physical Sciences, Indian Institute of Science Education and Research (IISER) Kolkata. Mohanpur 741246, India.}
	\address{$^2$Department of Condensed Matter Physics and Material Science, Tata Institute of Fundamental Research, Mumbai 400005, India.}
	\address{$^3$Center of Excellence in Space Sciences India, Indian Institute of Science Education and Research (IISER) Kolkata. Mohanpur 741246, India.}
	\address{$^4$Tata Centre for Interdisciplinary Sciences, TIFRH, Hyderabad 500107, India}
	\address{$^5$School of Physics, University of Hyderabad, Hyderabad 500046, India}
	
	\eads{\mailto{$^\dagger$sdghyderabad@gmail.com}}
	
	\begin{abstract}
		The interaction of light beams with resonant structures has led to the development of various optical platforms for sensing, particle manipulation, and strong light-matter interaction. In the current study, we investigate the manifestations of the bound states in continuum (BIC) on the in plane and out of plane shifts (referred to as Goos-H\"anchen (GH) and Imbert-Fedorov (IF) shifts, respectively) of a finite beam with specific polarization incident at an arbitrary angle. Based on the angular spectrum decomposition, we develop a generic formalism for understanding the interaction of the finite beam with an arbitrary stratified medium with isotropic and homogeneous components.  it is applied to the case of a Gaussian beam with $p$ and circularly polarized light  incident on a symmetric structure containing two polar dielectric layers separated by a spacer layer. For $p$-polarized plane wave incidence one of the coupled Berreman modes of the structure was recently shown to evolve to the bound state with infinite localization and diverging quality factor coexisting with the other mode with large radiation leakage (Remesh et al. Optics Communications, 498:127223, 2021). A small deviation from the ideal BIC resonance still offers resonances with very high quality factors and these are exploited in this study to report giant GH shifts. A notable enhancement in the IF shift for circularly polarized light is also shown. Moreover, the reflected beam is shown to undergo distortion leading to a satellite spot. The origin of such a splitting of the reflected beam is traced to a destructive interference due to the left and right halves of the corresponding spectra.
	\end{abstract}
	
	\vspace{2pc}
	

	\section{\label{sec:level1}Introduction}
	
	It is now well understood that in contrast to the predictions of geometrical optics, a finite beam can exhibit both spatial and angular shifts under reflection and transmission \cite{Bliokh2013}. The in-plane and out of plane spatial shifts, customarily known as the  Goos–H\"{a}nchen (GH) and Imbert–Fedorov (IF) shifts, have been studied in great detail, not only for their fundamental interest, but also for their numerous potential applications. While GH shift is sensitive to the pole of the reflection/transmission coefficients, it may lead to different amount of shifts for $p$-{(TM or Transverse Magnetic)} and $s$-{(TE or Transverse Electric)} polarized plane waves. In contrast, the IF shift owes its origin to the angular momentum carried by the incident beam (for example, for circularly polarized plane waves) leading to the so-called spin-Hall effect of light \cite{Bliokh:2015,Zhang2013,Kumar2022}. In case if the light is structured (as in case of evanescent waves or beams), it can lead to many other manifestations of spin-orbit interaction (SOI) like transverse spin, elusive Belinfante momentum, spin locking etc (see \cite{SDG2019}, along with \cite{Bliokh2013,Bliokh:2015} and the references therein). {Recently, the Optical Spin Hall effect (spin-orbit locking) has been experimentally demonstrated using a $\epsilon$-negative/$\mu$-negative \cite{Chen2017} and a hyperbolic waveguide \cite{Hong2021}}. Earlier studies on the spatial shifts were carried out with dielectric or metal-dielectric layered media, where such structures facilitated the excitation of guided or surface modes with very high local fields and strong confinement. As a consequence the otherwise tiny shifts (order of wavelength or less) could be enhanced significantly exploiting the large dispersion around these resonances \cite{Chen2019,Namdar2013, SDG2012}. . Recent studies focus on lower dimensional systems like  metasurfaces or 2D materials like doped graphene \cite{Zhang2013,Wen2017,Marini:2018}. Pump controlled atomic medium has been shown to lead to tunable GH shifts \cite{Wan2020}. There have been numerous applications for new sensing technology, precision measurements \cite{Jiang2017,Hesselink2006,He2008,Dadoenkova_2017}, optical switching \cite{Shimokawa2000} to name a few. Another interesting direction has opened up exploiting the weak value amplification techniques to amplify the weak shifts \cite{Hosten2008,Athira2020,Goswami2016,Athira2022b}. 
	\par 
	In the context of extra narrow resonances, the bound states in continuum (BIC) have lot of promises, since they offer infinite confinement with diverging quality factors. In view of the exotic fundamental physics and potential for novel applications they have been studied extensively resulting in vast literature \cite{DouglasStone2016, azzam2021}. Though originally proposed for a quantum mechanical model \cite{Neuman1929}, studies on BIC now extend to various different branches of physics with typical classifications like symmetry protected, accidental, Friedrich-Wintgen etc. \cite{DouglasStone2016,Friedrich1985,Koshelev:20,Monticone2018, Gopal:2021}. The exotic properties of BICs such as the high Q-factor and field localization have been exploited for different ends, such as allowing them to act as efficient lasing cavities \cite{Hirose2014,Miyai2006}. In addition, the topological nature of BICs found in photonic crystals allows them to emit light carrying orbital angular momentum \cite{Zhen2014,Iwahashi:11}. They can also be used in sensing, using the shift in the resonances with the change in the background refractive index. The strong dependence of the resonance position on the refractive index of the background medium of a BIC supporting structure can be used as a non-destructive  sensing technique to measure the refractive index in biological systems \cite{Romano2018}. Very recently BIC has been exploited to show large enhancements in the in-plane GH shifts  and its possible application for temperature and refractive index sensing \cite{wu2019,Zheng2021}. However, both these studies use Artmann formula \cite {artmann} for calculating the GH shifts. It is now well understood that the Artmann formula hinges on the stationary phase approximation, which does not hold for abrupt variation of phase near the extra high-Q resonances \cite{lai2086,GOLLA2011}. It is thus necessary to adopt a more rigorous theory based on angular spectrum decomposition to calculate the shifts for BIC.
	\par
	One of the simplest ways to implement a BIC in photonic systems is to make use of the  Fabry-Perot type BIC, where  light is trapped between two resonant structures, which act as perfect reflectors \cite{Remesh2021,Li2016,sakotic2020,Borisov2008}. In this context metallic  and polar dielectric films are attractive since at specified frequencies they can exhibit epsilon near zero (ENZ) behavior \cite{Li2016,Chen2013,Vassant2012a}. {Polar dielectrics materials are characterized by its longitudinal optical (LO) and transverse optical (TO) phonon frequencies $\omega_L$ and $\omega_T$ respectively. The real part of the dielectric permittivity of these materials tend to zero near $\omega_L$.} Very thin polar dielectric films have an additional advantage that they support the Berreman modes \cite {Vassant2012a,Berreman1963,Passler2019}. {These modes can be interpreted as virtual modes that emerge near the LO phonon frequency inside the light-cone for thin polar dielectric films. The absorption peak observed at the LO phonon frequency for thin films \cite{Berreman1963} can be attributed to the excitation of these modes \cite{Vassant2012a}}.  Being inside the lightcone, these leaky modes can be excited using propagating waves. The strong field confinement upon excitation of these modes has lead to several interesting nonlinear optical applications, for example, second-harmonic generation \cite{Passler2019,Aleksei2023}. In a recent paper a symmetric structure with two polar dielectric films separated by a air gap supporting coupled Berreman modes was studied. {  The Berreman modes in each film can couple to form a symmetric and antisymmetric mode, one of which can have null-width for suitable set of parameters, implying the transition to BIC} \cite{Remesh2021}. It was shown that the ENZ condition was essential for the BIC, since as mentioned above the polar dielectric layers then behave as ideal reflectors (in absence of losses).
	\par
	It is quite suggestive that the high quality factor, field localization and dispersion of BICs would make them an ideal candidate for enhancing the beamshifts. Even the quasi BICs (occurring slightly away from the ideal BIC resonance) offer enhanced quality factors. In this paper, we  study the beam profiles and shifts of a Gaussian beam incident on the  structure supporting coupled Berreman modes described above. While the previous study determined the effects of a $p$-polarized plane wave, our current formalism is suited to a generic  beam at arbitrary angle of incidence. In developing the formalism and the code we followed \cite{Bliokh2013} adding few important improvements. The formalism is applicable to general beams such as vector beams or OAM-carrying beams, though the results for only linearly or circularly polarized Gaussian beams are presented in this paper. We show an enhancement of the shift due to the strong field enhancement caused by quasi-BIC. We report a giant enhancement in the GH shift for incident $p$-polarized beam, while there is a discernible enhancement in the IF shift for circularly polarized light. We also report a distinct deformation of the reflected beam resulting in the formation of a satellite spot which is explained by means of an intriguing destructive interference.
	\par
	The organisation of the paper is as follows. After a summary of the general formulation of the problem, we describe our system supporting the coupled Berreman modes in Section \ref{Section2}. In Section \ref{Section3}, we discuss the numerical results as regards the beam profiles both in the momentum and space domains for both reflected and transmitted beams. The beamshifts are then calculated from the spatial profiles of the reflected and transmitted beams. Finally we summarize the main results and future outlook in Conclusions.
	


	\begin{figure}
		\centering
		\includegraphics[width=0.7\textwidth]{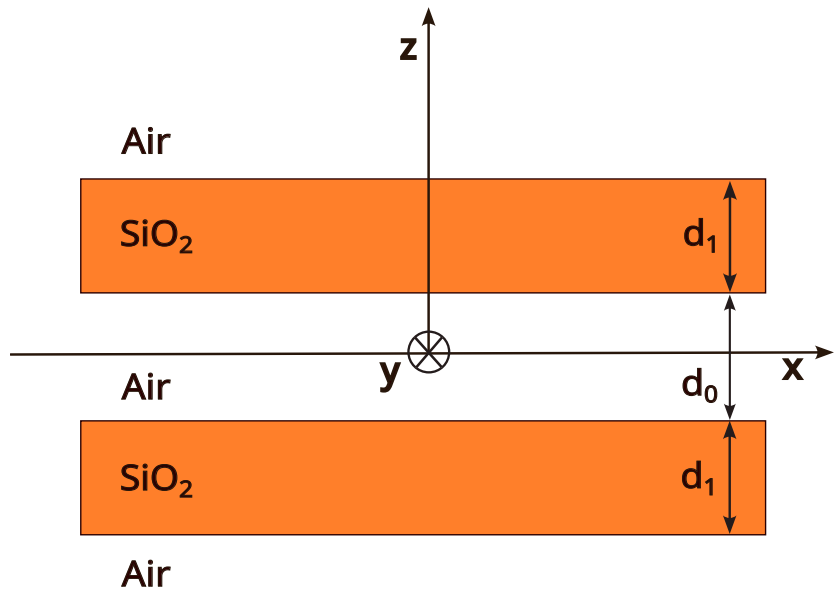}
		\caption{Schematics of the structure  comprising of two identical SiO$_2$ films of thickness d$_1 = 0.02\mu$m  each separated by a spacer layer of air with thickness $d_0 = 5.6\mu$m.}
		\label{fig:structure}
	\end{figure}
	
	\section{Formulation of the Problem}\label{Section2}
	In this Section we outline the procedure for a generic code for beam shape and associated shifts of the reflected and transmitted beams for a general multilayered medium for incidence of a vector beam at an arbitrary angle of incidence. Each constituent of the multilayered medium is assumed to be isotropic and homogeneous. The code can handle any beam ranging from a simple Gaussian to more complicated LG, or radially / azimuthally polarized vector beams. Our procedure is based on the approach developed by Bliokh et al. \cite{Bliokh2013} albeit for a single interface, except for two notable improvements. The first one concerns the strong paraxial approximation allowing for linear deviations for the wave vector deflections from the central wave vector in the beam frame (see eq. (3.9) in \cite{Bliokh2013}). Note that this approximation renders the theory not applicable to normal incidence, since for normal incidence lowest order contributions come from quadratic and higher order terms. Though we deal with paraxial beams using a Jones formalism we do not use any approximations for the polar and azimuthal angles for any off-axis $k$ vector component. The second improvement concerns the use of exact values of complex amplitude reflection and transmission coefficients for the structure as opposed to the  Taylor series expansion of these quantities retaining only the linear terms as in \cite{Bliokh2013}. Evidently, both these improvements make it impossible to have closed form expressions for the beam shifts, which is one of the major achievements of reference \cite{Bliokh2013}. Another limitation of the current approach stems from the Jones formalism with no reference to the longitudinal field components. One can not thus study transverse spin effects which draw their origin from the longitudinal component in tight focusing. We have tested our code for both the cases of normal and oblique incidence and reproduced the results of references \cite{She:2015,Rizza:2017} and \cite{Bliokh2013}, respectively. Note that we are empowered to generate the output beam profile which was missing in \cite{She:2015,Rizza:2017} to explicitly confirm the predictions of their study. We now list out the major steps retaining the notations of \cite{Bliokh2013} wherever possible and later apply the method to study our system supporting BIC. Here we use only incident Gaussian beam, though the formalism can yield results for other more general beams.
	\par
	Consider a Gaussian beam having the normalized spectrum given by \cite{Bliokh2013,Zhu2021}
	\begin{eqnarray}\label{gaussian}
		\ket{\mathbf{E}_i} = \frac{w_0}{\sqrt{2\pi}}\exp(-(k_x^2+k_y^2)w_0^2/4) (A_p\hat{e}_p + A_s\hat{e}_s), \label{eq:gaussian_beam}
	\end{eqnarray}
	be incident on the stratified medium. In eq. \ref{eq:gaussian_beam}, $w_0$ is the beam width and the second parenthesis represents the state of polarization of the beam with subscripts $p,s$ labelling the mutually orthogonal $p$ and $s$ polarized components. Let the central wave vector of the beam be incident at an angle $\vartheta_i$ (with corresponding reflection and transmission angles $\vartheta_r$ and $\vartheta_t$, respectively). For any off-axis wavevector component $(k_x,k_y)$ in the beam frame, the local polar and azimuthal angles, for example, for the incident beam are given by \cite{Zhu2021} 
	\begin{eqnarray}\label{theta}
		\theta_i = \tan^{-1} \left( \frac{\sqrt{k_y^2 + (k_x\cos \vartheta_i + k_z\sin\vartheta_i)^2}}{-k_x\sin\vartheta_i + k_z\cos\vartheta_i} \right),
	\end{eqnarray}
	\begin{eqnarray}\label{phi}
		\phi_i = \tan^{-1}\frac{k_y}{k_x \cos\vartheta_i+k_z\sin\vartheta_i},
	\end{eqnarray}
	where $k_z=\sqrt{1-k_x^2-k_y^2}$. Such local angles $\theta_a$ and $\phi_a,\,a = r,t$ are obtained also for reflected and transmitted beams and facilitate the rotational transformation from the beam frame to the basis of $p$ and $s$ modes. In order to calculate the output spectra of the reflected and transmitted beams one would require the corresponding reflection and transmission coefficients for each spatial harmonic for both $p$ and $s$ polarizations. These quantities ($a_p, a_s$, respectively) are calculated using the characteristic matrix approach \cite{sdgbook} for each of these harmonics. Thus the output spectra  are given by \cite{Bliokh2013}
	\begin{equation}
		\ket{\mathbf{E}_a} = U^\dag_a F_a U_i  \ket{\mathbf{E}_i}, \label{eq:4}
	\end{equation}
	where
	\begin{eqnarray}
		U_a = \hat{R}_y(\theta_a) \hat{R}_z(\phi_a) \hat{R}_y(-\vartheta_a),\label{eq:5}\\
		F_a = \mbox{diag}(a_p, a_s) \label{eq:6}, a=r,t,
	\end{eqnarray}
	with $i, r,t$, denoting the relevant quantities for incident, reflected and transmitted fields. $R_{x,y,z}$ are rotation matrices about $x,y,z$ axes in the laboratory frame. $U_i$ in eq.\ref{eq:4} is obtained by setting $a=i$ in eq.\ref{eq:5}. As mentioned earlier Bliokh et al. used a linear approximation for calculating the local angles $\theta_a$ and $\phi_a$, and to approximate the reflection and transmission coefficients for off-central $k$ components, which is applicable only in the strong paraxial regime. We calculate them exactly using eq. \ref{theta} and eq. \ref{phi} allowing us to employ beams with larger transverse $k$ spread. As mentioned above the matrix $F_a$ contains the information for reflection and transmission properties of the structure. Feeding the information for Fresnel coefficients and rotation matrices for each wavevector (on or off-central) of the incident Gaussian beam, the reflected and transmitted spectra are calculated. The inverse {Fourier} transform of the spectra given by eq. \ref{eq:4} then yields the corresponding beam spatial profiles. In order to investigate the spatial in-plane (Goos-Hänchen) and out-of-plane (Imbert-Fedorov) shifts, we calculate the centre of gravity of the reflected and transmitted beams using \cite{Remesh2022}
	\begin{eqnarray}\label{cg}
		x_0^{a} = \frac{\int x|\vec{E_a}|^2 \,dx \,dy}{\int |\vec{E_a}|^2 \,dx \,dy},\\
		y_0^{a} = \frac{\int y|\vec{E_a}|^2 \,dx \,dy}{\int |\vec{E_a}|^2 \,dx \,dy},
	\end{eqnarray} 
	where $a=r,t$ refer to the reflected and transmitted beams, respectively.
	
	\par
	We now apply the above formalism to the symmetric multilayered medium in Fig. \ref{fig:structure}. Recall that a single  SiO$_2$ film on a gold substrate was studied \cite{Vassant2012a} for  probing the leaky Berreman modes \cite{Berreman1963}. One of the coupled Berreman modes of the structure shown in Fig. \ref{fig:structure} was shown to evolve to the BIC under continuous tuning of the system parameters \cite{Remesh2021}. The dielectric function $\epsilon_d$ of SiO$_2$ was taken to be \cite{Vassant2012a},
	\begin{eqnarray}\label{sio2}
		\epsilon_{d} = \epsilon_\infty\frac{\omega^2-\omega_L^2+i\omega\Gamma}{\omega^2-\omega_T^2+i\omega\Gamma}.
	\end{eqnarray}
	Typical parameters for polar  SiO$_2$  are $\epsilon_\infty=2.095$, $\omega_L=2.30\times10^{14}$ rad/s, $\omega_T=1.98\times10^{14}$ rad/s, $\Gamma=1.35\times10^{13}$ rad/s. However, in order to look for BIC, losses were ignored by setting $\Gamma=0$ \cite{Remesh2021}. It was shown earlier that Epsilon-near-zero (ENZ) behavior at $\omega=\omega_L$ is essential for the occurrence of BIC, since under ENZ condition the polar dielectric layers act as perfect mirrors. The angle $\theta$ at which the BIC occurs can be found from the phase-matching condition, which is given by, 
	\begin{eqnarray}\label{phase_match}
		k_0d_0\cos \theta = n\pi,
	\end{eqnarray}
	where $k_0$ is the wavevector of light in vacuum, $d_0$ is the thickness of the spacer layer of the structure and $n$ is an integer. The angle of incidence of the central wavevector of the incident Gaussian beam was chosen near this BIC angle, which would hence excite the quasi-BIC resonances with finite widths in the structure.
	\begin{figure}
		\centering
		\includegraphics[width=0.7\textwidth]{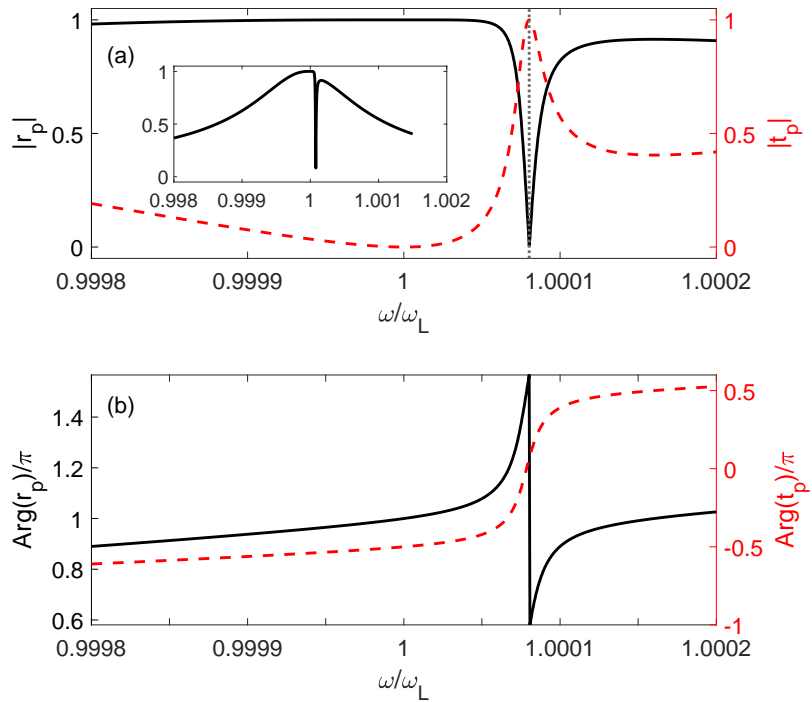}
		\caption{(a) Absolute value of reflection (solid black) and transmission (red dashed) coefficients and (b) the corresponding phases of reflection (solid black) and transmission (red dashed), respectively,  for a $p$-polarized plane wave at a fixed angle of incidence $ \vartheta_i = 47^{\circ}$. The inset in the top panel shows the trend over a broader range of frequencies. The vertical dotted line in (a) marks the location of quasi-BIC.}
		\label{fig:randt}
	\end{figure}

	\begin{figure}
		\centering
		\includegraphics[width=0.7\textwidth]{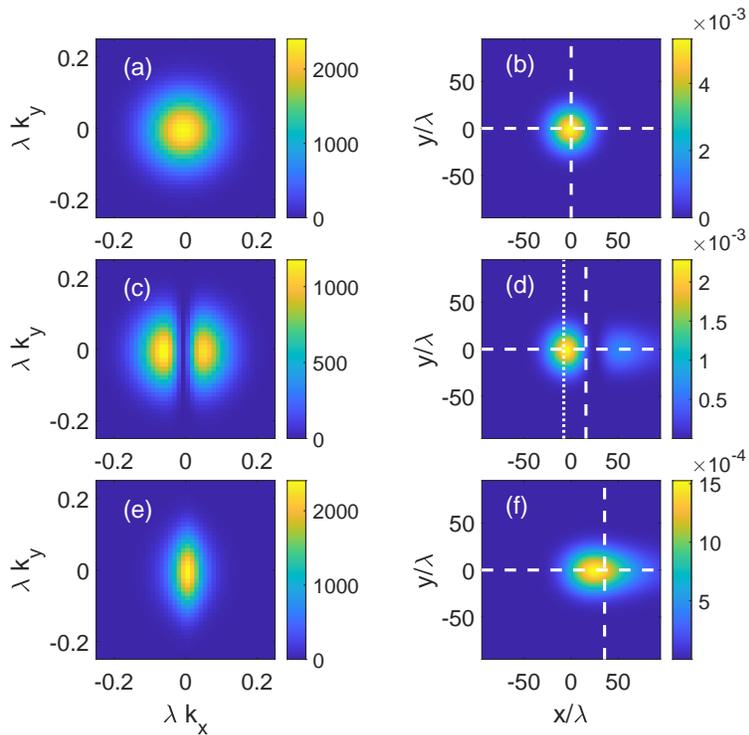}
		\caption{Intensity spectra of the (a) incident, (c) reflected and (e) transmitted beams, respectively for $p$-polarized Gaussian beam with a beam waist of $w_0=15\lambda$. The corresponding spatial intensity distributions are shown in the right panels (b), (d) and (f), respectively. The white-dashed cross hair shows the center of gravity of the beam. The intersection between the white dotted line and dashed line in (d) shows the CG after suppressing the faint right spot.}
		\label{fig:beam}
	\end{figure}
	\section{Results and Discussion}\label{Section3}
	In this Section we present the numerical results. We proceed as follows. We recall some of the relevant results from the earlier study  \cite{Remesh2021} pertaining to the reflection and transmission coefficients for $p$-polarized plane wave incidence just to focus on the particular quasi BIC at a given angle of incidence. Recall that as predicted by eq. \ref{phase_match}, there can be higher order BICs discussed in detail in \cite{Remesh2021}. An angle close to this is then chosen to be the angle of incidence of the central wave vector of the $p$-polarized Gaussian beam. We have also looked at circularly polarized Gaussian beams to infer on the IF shifts. Henceforth, the  formalism developed in the previous section is applied to obtain the spectra, their originals and the corresponding in-plane and out of plane shifts.
	\par
	We recall the parameters used for the numerical studies. The relevant parameters for eq. \ref{sio2} are $\epsilon_\infty=2.095$, $\omega_L=2.30\times10^{14}$ rad/s, $\omega_T=1.98\times10^{14}$ rad/s, $\Gamma=0$ rad/s. The thickness of the SiO$_2$ and air layers were taken to be $d_1 = 0.02 \,\mu $m and $d_0 = 5.6\,\mu$m. The angle of incidence of the beam was chosen to be $\vartheta_i = 47\degree$, which is close to the angle at which BIC occurs (one of the resonances $n= 1$ of eq. \ref{phase_match}).
	Recall that BIC requires the ENZ condition of $\omega/\omega_L=1$.
	\par
	The results for $p$-polarized plane wave incidence are shown in Fig. \ref{fig:randt}, where we have plotted the absolute values of the amplitude reflection $r_p$ and transmission $t_p$ coefficients along with their phases as  functions of frequency at a fixed incidence angle of $\vartheta_i=47 \degree$. Since we are working at angles slightly away from the BIC, the quasi-BICs are excited at a frequency shifted from $\omega=\omega_L$. 
	One can notice the sharp dip (peak) in the reflection (transmission) coefficients in Fig. \ref{fig:randt} at $\omega/\omega_L=1.0000803$. The inset in Fig. \ref{fig:randt}a shows the reflection coefficient over larger frequency domain in order to appreciate the ultra-narrow BIC dip. It is important to note the phase jump in reflection at the BIC dip, which, as will be shown below, plays an important role in determining the reflected beam profile.
	\par
	In the context of the $p$ polarized Gaussian beam incidence we choose the frequency and angle of incidence corresponding to the quasi-BIC dip location in \ref{fig:randt}(a), namely, $\omega=1.0000803 ~\omega_L$ and $\vartheta_i = 47\degree$, so that a spatial harmonic (plane wave) corresponding to the central wavevector of the beam would excite the quasi-BIC. The beam waist of the incident Gaussian beam was chosen to be $w_0 = 15\lambda$, where $\lambda= 2\pi c/\omega$ is the wavelength of the light used.
	\begin{figure}[t]
		\centering
		\subfloat{
			\includegraphics[width=0.49\columnwidth]{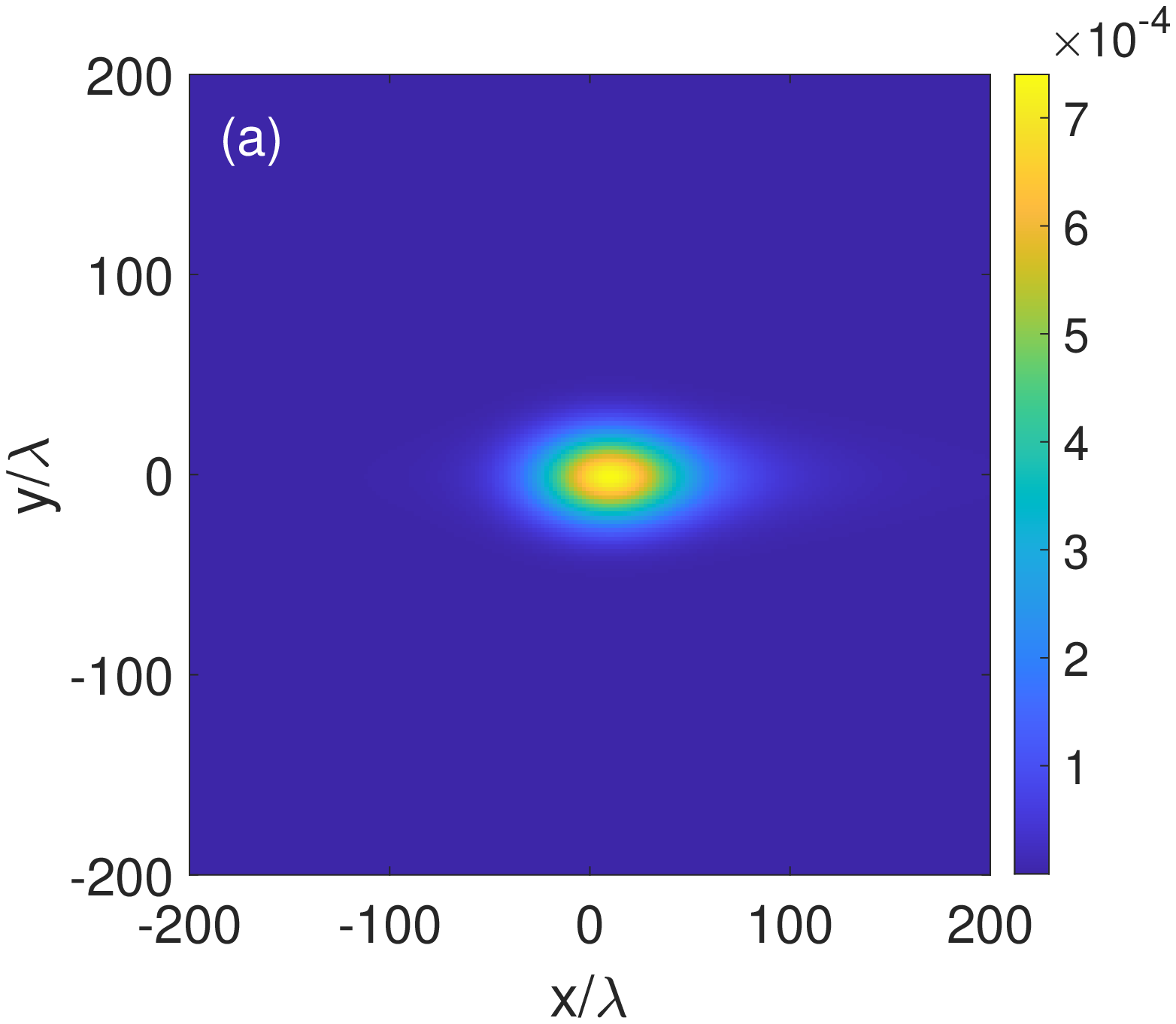}
		}\hspace*{\fill}%
		\subfloat{
			\includegraphics[width=0.49
			\columnwidth]{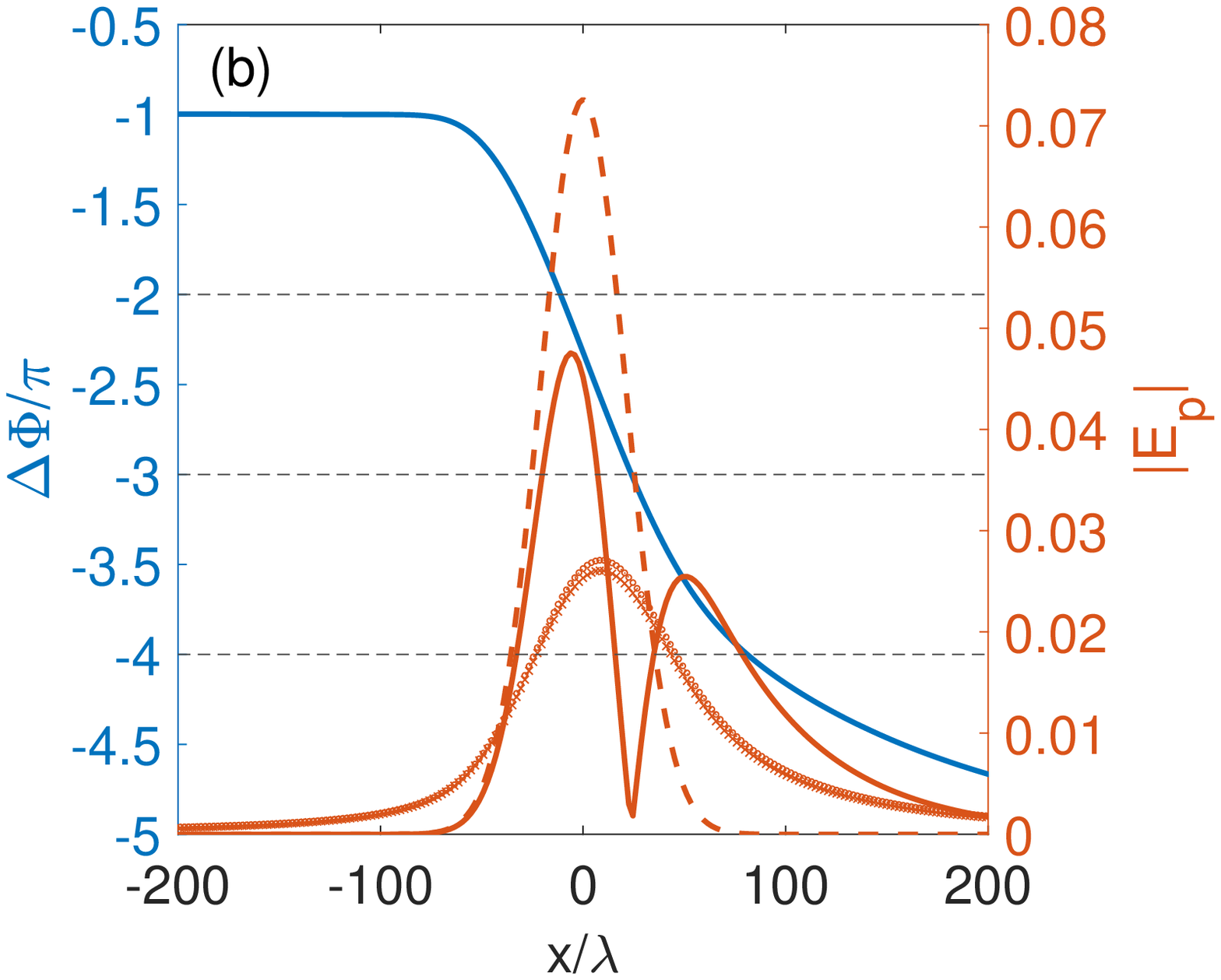}
		}
		\caption{(a): Spatial intensity profile of the reflected $p$-polarized light after suppressing the right lobe in Fig.\ref{fig:beam}c. (b): Phase difference of the spatial field distributions $\Delta \Phi$ between the right and left lobes as a function of $x$ for $y=0$. Right axis: Absolute value of the $p$-polarized  incident electric field amplitude (Dashed line), reflected $p$-polarized light (Solid line), and reflected $p$-polarized light after suppressing the left ({circle})/right ({cross}) lobe as a function of $x$ for $y=0$.} \label{Fig:4}
	\end{figure}
	\par
	The intensity spectra of the incident, reflected and transmitted beams (obtained using eq. \ref{eq:gaussian_beam} and eq. \ref{eq:4} are shown in Figs. \ref{fig:beam}a,c,e. The corresponding spatial intensity distributions (from inverse {Fourier} transforms) are shown in the right panels. Clearly, the two lobes (hereafter referred to as left and right lobe) in Fig. \ref{fig:beam}c owe their origin to the sharp BIC dip in reflection for the central component  with analogous dips for nearby spatial harmonics wherever the angle of incidence is near the quasi-BIC angle. Note that $k_x=k_y = 0$ corresponds to the central wavevector of the obliquely incident Gaussian beam in the beam frame. The signature of BIC shows up as a bent vertical  strip in the phase plot in the ($k_x,k_y$) plane (not shown). One can also relate the origin of the dark strip to the $\pi$ phase change as one moves from the left to the right lobe. The inverse Fourier transforms of the momentum space spectra yield the corresponding spatial intensity distributions (see Figs.\ref{fig:beam}b,d,e). Even though the incident beam is $p$-polarized, the reflected and transmitted light spatial intensities have contributions from $s$-polarization also because of $p-s$ mixing for off-central wave vector components (albeit about five orders of magnitude smaller than the former). 
	Having obtained the reflected and transmitted spatial profiles, the GH and IF shifts of the center of gravity (CG) for relevant polarizations of the incident beam can now be calculated using eq. \ref{cg}. Due to the split spectrum of the reflected light, the corresponding spatial profile exhibits distortion leading to a faint additional spot as shown in Fig. \ref{fig:beam}d. We refer to the bright and lighter spots in Fig. \ref{fig:beam}d as the main and satellite spots. Somewhat reminiscent emergence of satellite spot has been reported in reflection of higher order Gaussian beams, albeit due to tight focusing \cite{SDG2012,GOLLA2011}. The origin of these spots can be traced to interesting destructive interference effects discussed below. Finally in order to assess the shift of the reflected beam, we calculate the same retaining both the spots and also by suppressing the satellite spot {by setting the field amplitude to be zero for $x/\lambda> 25$ (location of the intensity minimum of the beam) in the reflected beam profile in Fig. \ref{fig:beam}d}. The  white dashed cross-hair marks the CGs of the overall intensity distributions, while the intersection of the dashed and dotted line mark the CG with suppressed satellite spot.
	\begin{figure}[t]
		\centering
		\includegraphics[width=0.7\textwidth]{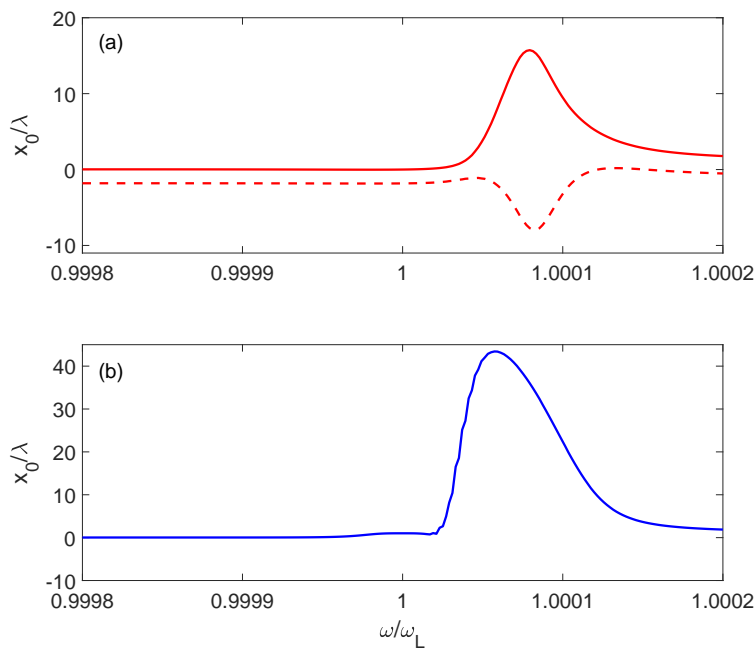}
		\caption{In plane Goos-Hänchen shift $x_0/\lambda$ for $p$-polarized Gaussian beam as a function of the normalized frequency at an angle of incidence $\vartheta_i=47 \degree$ for (a) reflected  and (b) transmitted  beams. The solid (dashed) line in (a) gives the Goos-Hänchen shift for the overall (satellite-suppressed) pattern.}
		\label{fig:gh}
	\end{figure}
	\par
	We now elaborate on the emergence of satellite spots in Fig.\ref{fig:beam}d. The reflected beam can be thought of as superposition of two spectra corresponding to the left and right lobes in momentum space (Fig.\ref{fig:beam}d). We look at the inverse Fourier transform (spatial image) of these $k$-space lobes. Recall that the reflected spectra has contributions from both $p$ and $s$-polarizations because of the aforementioned $p-s$ mixing of off-central wave vector components. However, since $p$-polarization plays the dominant role in BIC, we retain only the $p$-component of the electric field $E_p$. This is justified because $|E_s| \ll |E_p| $, as pointed out before. The spatial intensity distribution corresponding to the left lobe (see Fig.\ref{fig:beam}c), { obtained by inverse Fourier transforming the spectra after setting the field amplitude to zero for $k_x>0$}  is plotted in Fig.\ref{Fig:4}a. The spatial intensity distribution of the left and right lobes happens to be almost identical. However, corresponding phases play a very important role. In order to probe the underlying interference phenomena, we take a cross-section of this spatial image at $y=0$ and study the trend of amplitude and phase. The results are shown in Fig.\ref{Fig:4}b. As mentioned earlier, the amplitudes of the left and right lobes more or less match whereas the phase difference $\Delta\Phi$ between the two shows a monotonically decreasing behavior. At the point where $\Delta\Phi$ reaches an odd multiple of $\pi$ (here, $3\pi$) one can see a clear signature of destructive interference, causing the dip, resulting in the main and satellite spots, the left main being the stronger one.
	\par
	In what follows we present the explicit results for the enhancement of beamshifts due to the quasi-BIC. For studying Goos-H{\"a}nchen shift, we consider $p$-polarized incident light. Analogous calculations can be carried out for $s$-polarized light also, but the coupled Berreman modes near the ENZ condition, whose mutual interactions leads to the BIC cannot be efficiently excited for $s$-polarized incidence \cite{Remesh2021}. In Fig. \ref{fig:gh}a, the solid (dashed) line gives the shift for the overall (main) reflected spot as a function of normalized frequency, which clearly shows the enhanced shift upon excitation of the quasi-BIC. Note that assessing the GH shift of the distorted reflected beam (overall spatial distribution retainng both main and the satellite spots) can be quite misleading because of the loss of its Gaussian character of the reflected beam. In contrast, the shift retaining only the main spot (suppressing the satellite in Fig.\ref{fig:beam}d), in our opinion, gives a better estimate of the acceptable GH shift. Importantly these two approaches yield results with opposite signs. Needless to mention that the Imbert-Fedorov shift is negligible for $p$-polarization as expected. The corresponding shift for the transmitted beam is shown in Fig. \ref{fig:gh}b. Clearly the GH-shifts for both reflected and transmitted beams receive giant enhancements thanks to the large quality factor of the quasi-BIC. {Changing the structural parameters of our system can change the angle of incidence or quasi-BIC frequency. The different possible parameters in the available parameter space might lead to different beam shifts, different relative intensities of the satellite spots, etc. However, the general features shown in our system (enhancement of shifts, distortion of reflected beam and splitting into unequal spots, etc) are seen throughout the parameter space we explored.}
	\begin{figure}[t]
		\centering
		\subfloat{
			\includegraphics[width=0.49\columnwidth]{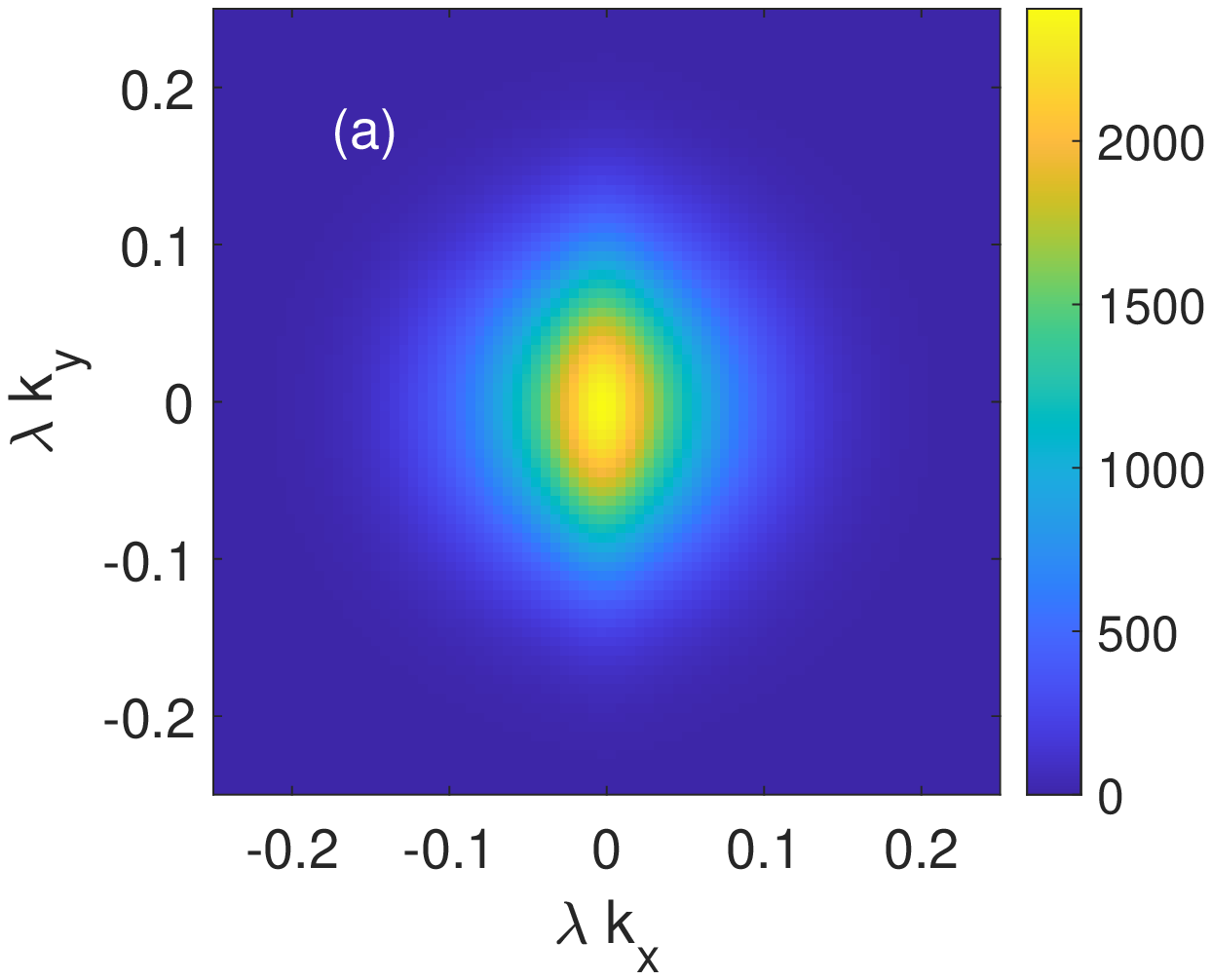}
		}\hspace*{\fill}%
		\subfloat{
			\includegraphics[width=0.49
			\columnwidth]{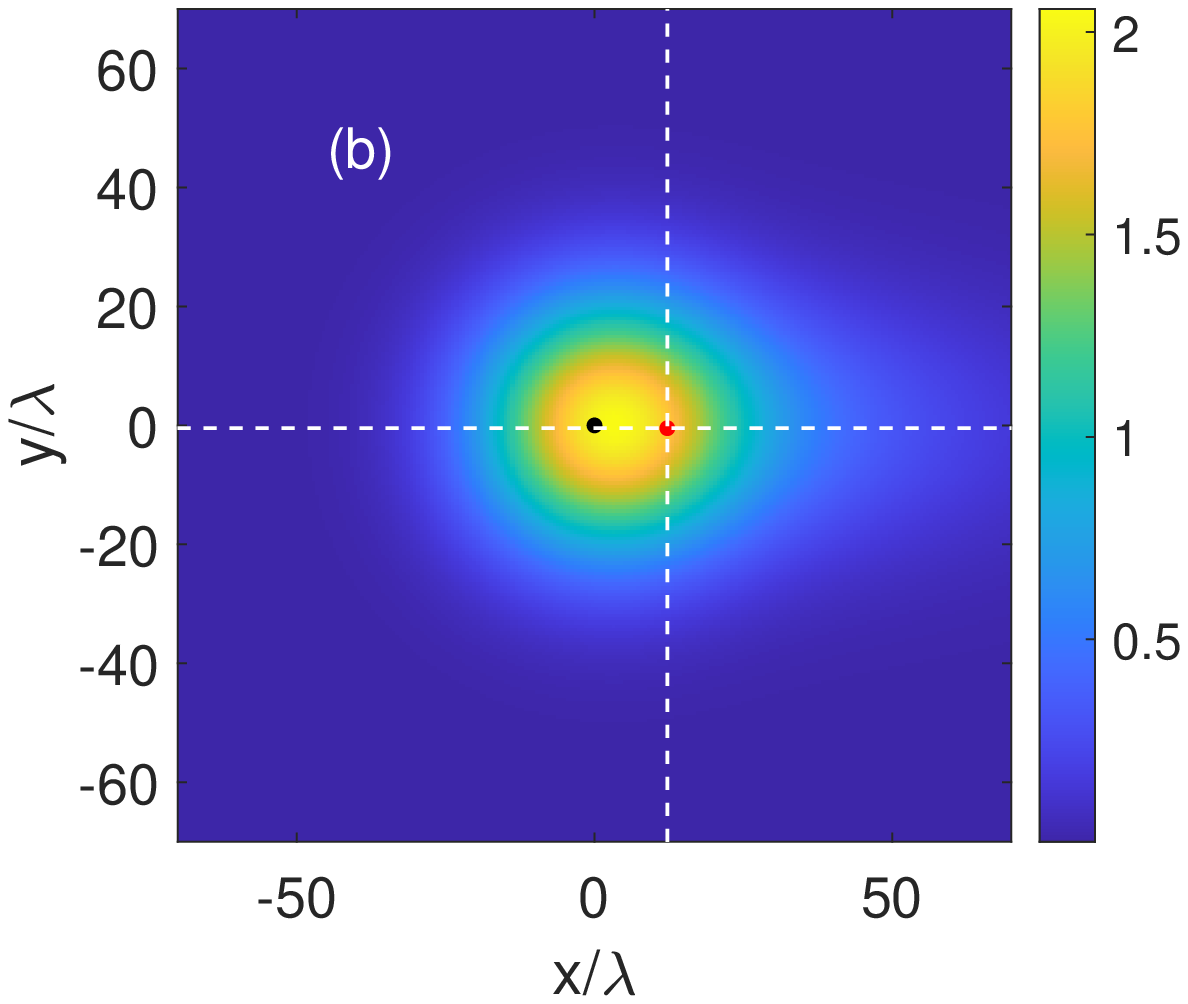}
		}
		\caption{(a) Transmitted beam intensity spectrum for incident left circularly polarized Gaussian
			beam. (b) Its corresponding spatial intensity distribution. The black (red) dot gives the center of gravity
			of the incident (transmitted) beam. Notice the small shift in the $-y$ direction.} \label{Fig:6}
	\end{figure}
	
	\par
\begin{figure}[t]
		\centering
		\includegraphics[width=0.7\textwidth]{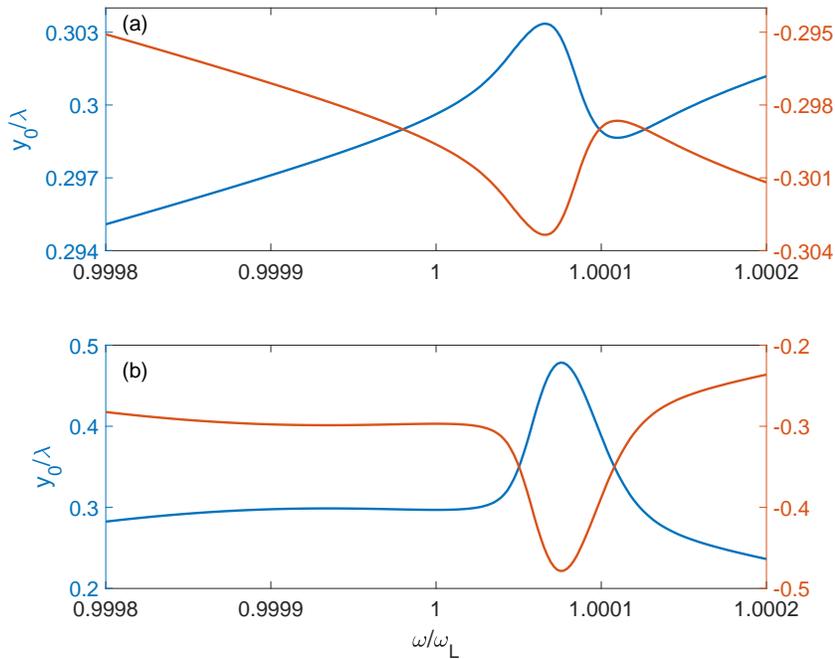}
		\caption{Imbert-Fedorov shift for left-circular (blue) and right-circular (orange) polarized Gaussian beams at various frequencies. Trend for reflection is plotted in (a) and transmission in (b).}
		\label{fig:if}
	\end{figure}

	Finally, we look at the out-of-plane IF shift for incident LCP light. The results for spectra and the corresponding spatial profiles for the transmitted beam are shown in Fig. \ref{Fig:6}. Clearly the transmitted beam retains the Gaussian character. The reciprocal behavior of the $k$ and $x$ domain is clearly seen from a comparison of Figs. \ref{Fig:6}a and b, since the narrower the spectra in $k_x$ direction, broader the spatial profile in the corresponding direction. As can be seen, the IF shift of the transmitted light is also associated with the in-plane GH shift. The similar but smaller enhancement of Imbert-Fedorov shift is observed for circularly-polarized light in Fig. \ref{fig:if}. The small shift is probably related to the paraxially incident beam that we have used. A highly focused non-paraxial beam has been shown to show strong SOI effects: for example, beams tightly focused by high-numerical-aperture lenses \cite{Kumar2022,Pal2020,Roy2022} or scattered by small particles \cite {Bliokh:2015,Soni2013}. Moreover our system supports BIC only for $p$- polarized light, since $s$-polarized light can not excite the Berreman modes. Since the circularly (or elliptically) polarized light is a prerequisite for IF shift it does not experience the giant shift as experienced by the lateral GH shift. Obviously, the magnitude of shifts is the same in LCP and RCP beams but in opposite directions.

	\section{\label{sec:level4}Conclusions}		
	In conclusion, we have studied a symmetric layered medium comprising of two polar dielectric Si$O_2$ films supporting bound states in continuum mediated by the coupled Berreman modes. We developed a general code for reflection and transmission of a Gaussian or vector beam through a stratified medium overcoming some of the strong paraxial approximations. The code is then applied to a Gaussian beam incident on the structure under study. We have shown theoretically that the extra-large quality factor and the resulting field enhancement associated with the quasi-BIC in the structure can lead to giant enhancements in the in-plane Goos-H\"anchen shifts which could be comparable to the beam spot size. Moreover, quasi-BIC is shown to distort the reflected beam to the extent of formation of a satellite spot. We systematically analyze the formation of the satellite spot as a consequence of an interesting destructive interference between the split lobes in the spectral profile of the reflected beam. We estimated the beam shifts for both linear and circular polarization of the incident Gaussian beam. Work is underway to consider richer structured beams (eg., LG or radially or azimuthally polarized vector beams) carrying angular momenta and will be reported elsewhere. Needless to mention that such studies are extremely important not only for fundamental understanding but also for novel nearfield applications for micro manipulation of particles.

	\ack
	SSB would like to acknowledge the support of Department of Science and Technology, Government of India. 
	\par 
	\noindent SSB and GR have contributed equally to this work.

	\section*{References}
	
	\bibliography{references}
	\clearpage
	
\end{document}